# A Quasigroup Based Cryptographic System

Maruti Venkat Kartik Satti

## 1. Introduction

This paper presents a quasigroup encryptor that has very good data-scrambling properties and, therefore, it has potential uses in symmetric cryptography. The purpose of the scrambler is to maximize the entropy at the output, even in cases where the input is constant. The great complexity associated with the task of finding the scrambling transformation ensures the effectiveness of the encryption process. Quasigroup (or Latin squares) encryption is a development that has permutation based scrambling [1-6] at its basis.

The output of the proposed encryptor is dependent upon the index numbers and the orders of the matrices (r, s) which are sent by the trusted authority. The encryption is also dependent on six multiplier elements that are generated by a secret algorithm based on the index numbers, the order of the matrices under consideration and nonce (random number generated by trusted authority). This *key* is updated by the network on a regular basis (once in every time interval that is far less that T, the time needed to use brute force to decrypt the key that is sent by the trusted authority). Figure 1 illustrates the quasigroup encryptor. The module has takes the raw data stream and randomizes it based on the encryption key (the encryption key), and the output data has desirable autocorrelation properties.

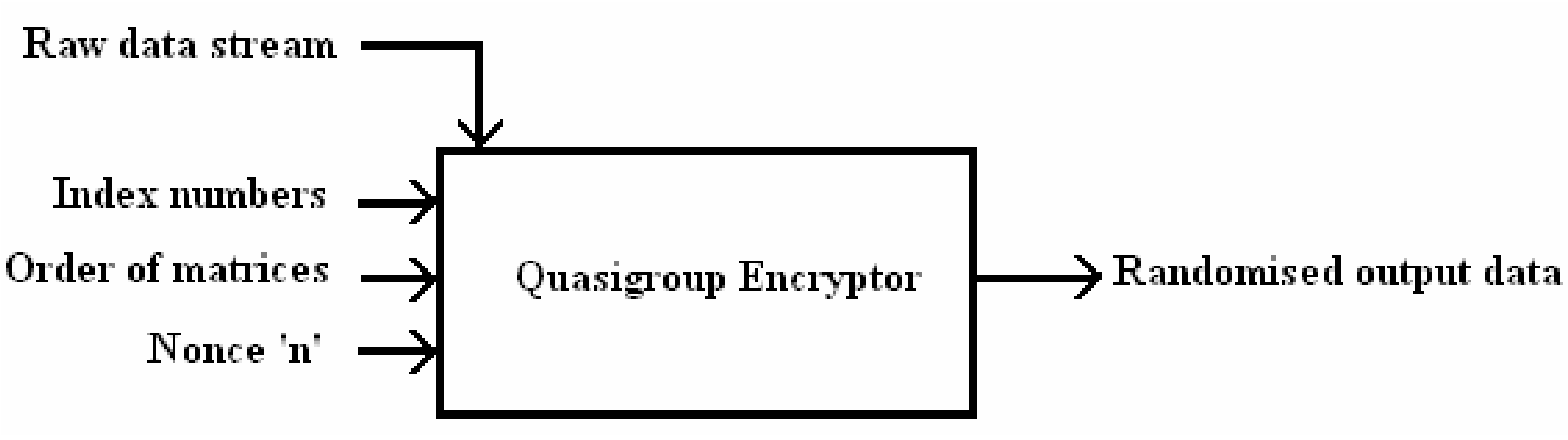

Figure1: Quasigroup encryptor

If we view the encryptor mainly from the perspective of maximization of entropy, it is appropriate to point out that quantum mechanical systems also may be used for such entropy transformation [7-15]. Entropy in a quantum system has a slightly different definition [7, 8], and because of the probabilistic nature of state reduction, uncontrolled randomness becomes an issue [9-11]. Nevertheless, one can devise quantum cryptographic systems that can be used in public-key cryptography applications [12-15], although it is not clear that they will be commercially successful in the near future. Because of the limitation of quantum systems related to practical implementation, permutation-based symmetric cryptographic transformations are going remain important. For earlier work on quasigroup cryptography, see [16-29]. Here I propose a new algorithm which distributes raw data stream based upon the indices and the order of the quasigroup under consideration. This unique key, which consists of the index numbers and the order *r* and *s* that are kept secret, is transmitted by the trusted authority in a way that the keys do not repeat till the characteristic network time expires.

## 2. Quasigroup definition and theory

A quasigroup G can be defined as a group of elements (1, 2, 3 … n) along with a multiplication operator such that for every element x and y there exists a unique solution such that z such that the following two conditions are obeyed:

$$x*a=z \qquad (1)$$
$$y*b=z$$

in the above defined equations the elements a, b and z belong to the Quasigroup 'G' . Or, a quasigroup is a binary system (Q,*) satisfying the two conditions

For any a, b belongs to Q there exists a unique x belongs to Q such that $a*x=b$ (2)
For any a, b belongs to Q there exists a unique y belongs to Q such that $y*a=b$.

Let us now consider the following example of a quasigroup G of order 4 for simplicity

| * | 1 | 2 | 3 | 4 |
|---|---|---|---|---|
| 1 | 2 | 3 | 1 | 4 |
| 2 | 4 | 1 | 3 | 2 |
| 3 | 3 | 4 | 2 | 1 |
| 4 | 1 | 2 | 4 | 3 |

A multiplication operator in a quasigroup in not typically like ordinary multiplication but it behaves as some sort of mapping between the row and the column indices. Let us suppose x=2 and a=3; the resulting z can be determined by looking up the element having the row index as 2 and the column index as 3, so the obtained value of z is 3 (z = G (2,3)).

## 3. Quasigroup Signal Processing

This section introduces the Multi-Level Indexed Quasigroup Encryptor (MLQE) works. First, let us define certain terms that are standard throughout the paper.

Input data: $d_1, d_2, d_3,\ldots d_n$ (3)
Output data: $e_1, e_2, e_3,\ldots e_n$
The two matrices: $R, S$
Multiplier Elements: $q_1, q_2, q_3,\ldots q_n$

The indices: $I_1, I_2, I_3, \ldots I_n$

The encryptor is defined by QE (stands for Quasi-Encryptor), and the decryptor is defined as QD (stands for Quasi-Decryptor).

## 3.1 Encryption:

Note that if Q is a quasigroup such that $a_1, a_2, a_3, \ldots a_n$ belong to it then the encryption operation QE, which is defined over the defined elements, maps those elements to another vector $b_1, b_2, b_3, \ldots b_n$ such that the elements of the resultant vector also belong to the same quasigroup.

Markovski and Dimitrova [29] show that mapping of an incoming stream of data depends on the initial multiplier element. I have chosen to index the isotopes of a particular quasigroup and use these indices to reference these matrices to the encryptor. In addition, our method encrypts the data with a multi ordered quasigroup.

The mathematical equation used for encryption (basic level [29]) is defined by:

$$E_a(a_1, a_2, a_3, \ldots, a_n) = b_1, b_2, b_3, \ldots b_n \qquad (4)$$

where the output sequence is defined by:

$$b_1 = a * a_1$$
$$b_i = b_{i-1} * a_i$$

where $i$ increments from 2 to the number of elements that have to be encrypted, and $a$ is the *hidden key* (*leader* in Markovski and Dimitrova terminology). Equation (4) describes a typical single level quasigroup encryptor.

Let us now attempt to describe the workings of equation (4) with the help of an illustration (Figure 2). We assume that the initial input data is given by the vector $a_1, a_2, a_3, a_4, a_5, a_6$. It is mapped to the vector $b_1, b_2, b_3, b_4, b_5, b_6$ by equation (4). The following steps are used during the process of encryption:

$$b_1 = a * a_1 = 2 * 2 = 1$$
$$b_2 = b_1 * a_2 = 1 * 4 = 1$$
$$b_3 = b_2 * a_3 = 4 * 1 = 4$$
$$b_4 = b_3 * a_4 = 4 * 2 = 5$$
$$b_5 = b_4 * a_5 = 5 * 3 = 1$$
$$b_6 = b_5 * a_6 = 1 * 3 = 2$$

The sequence thus obtained is given as an input to another level of the encryptor. This process is done for several times. Multiple levels of mapping ensure lower resemblance of the output data to that of the input data. It makes it harder for the eavesdropper to decrypt the data. In another implementation, the multiplier element is varied. In my implementation these multipliers are generated by a special algorithm called "MEG1" that generates the multiplier elements based on the index numbers, Nonce, r and s in this second implementation can be given by the following equations [29]:

$$E_{h1,h2,h2\ldots hn}\,(a_1,a_2,a_3\ldots a_n)=e_1,e_2,e_3,\ldots e_n \qquad (5)$$

where

$$e_1=a*a_1 \text{ and } e_i=e_{i-1}*a_i$$

In the above equation the incoming stream of data is first mapped using the first multiplier element *h1* then the resultant steam is mapped considering the second multiplier element *h2.* This process continues till all the multiplier elements are exhausted.

$$b_1=h_1*a_1;\ b_2=b_1*a_2;\ldots\ b_n=b_{n-1}*a_n \qquad (6)$$
$$c_1=h_2*b_1;\ c_2=c_1*b_2;\ldots\ c_n=c_{n-1}*b_n$$
$$.$$
$$.$$
$$e_1=h_n*s_1;\ e_2=e_1*s_2;\ldots e_n=e_{n-1}*s_n$$

where the vector $(h_1,h_2,h_3,\ldots h_n)$ consists of all the multiplier elements. In this approach, this encryption key is transmitted along with the quasigroup (this key is itself encapsulated by another layer of encryption). It is understood that in the above two approaches another reliable encryption algorithm is required to preserve the secrecy of the encryption. Furthermore, it is necessary to transmit the quasigroup that is being used for encryption, which is one of the main drawbacks of the above approach. Once the eavesdropper breaks the encapsulating cipher he has access to the quasigroup used for the encryption and all the other required information to get the data.

The third approach is the index based approach where the given data is encrypted through several levels the encryption as described in the next section. Let us see what happens in this changed order index based approach. Consider that this changed order encryptor is given the input of all 1s (as illustrated in Figure 3). We see that after the second level encryption the input vector is mapped to the sequence which has symbols ranging from 1 to the order of the second matrix. So, if we have an index key which references the matrices stored in the memory of the reception device, the eavesdropper would not know which matrix is stored at a given index. To further improve the efficiency of this quasigroup encryptor, we can include another function that arranges the quasigroups based upon the Nonce this makes the encryption more time dependent and it can be observed that at any given point of time the output of the encryptor is different even if the same set of indices are supplied to the algorithm.

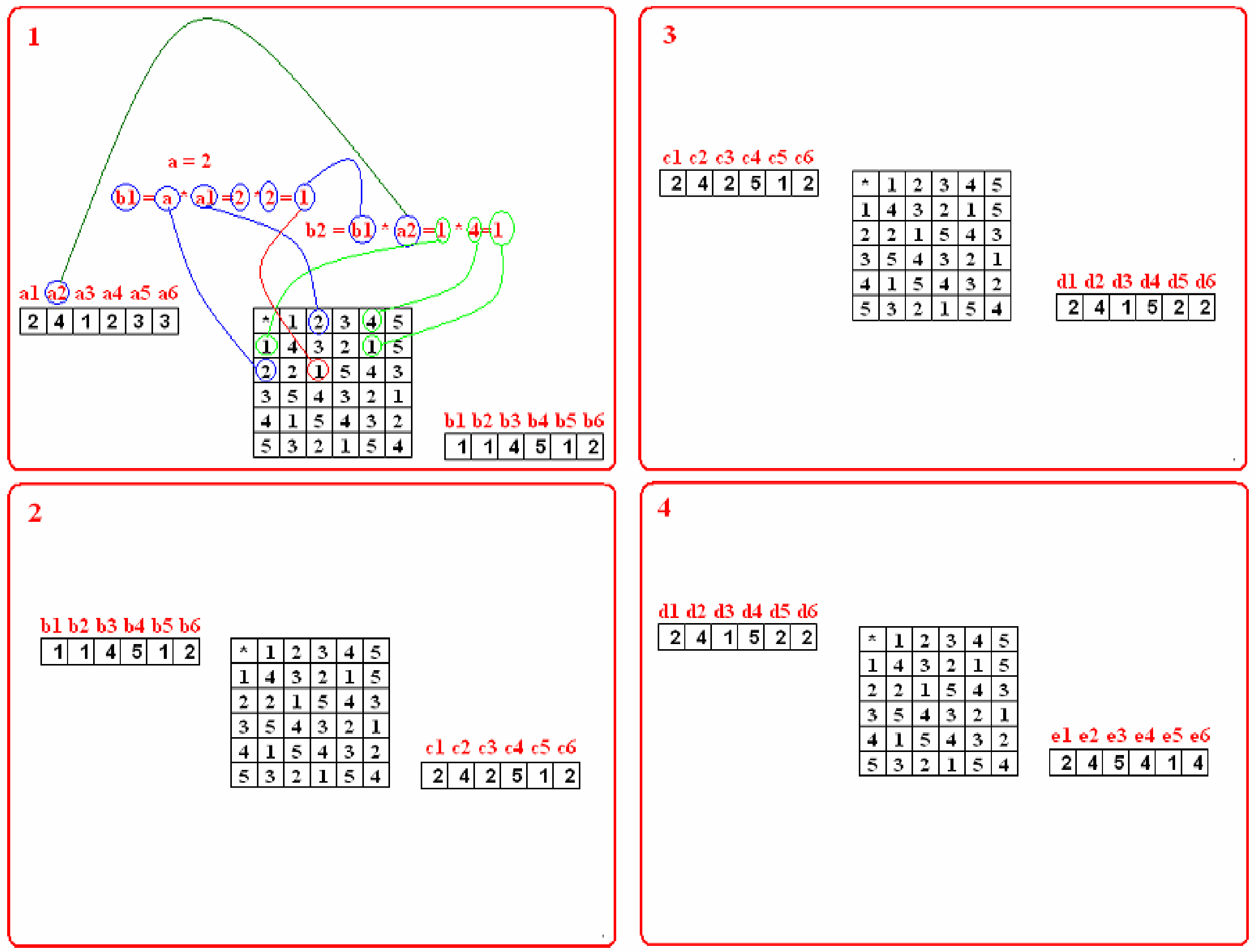

Figure 2: Quasigroup mapping using an order 5 quasigroup

| a1 | a2 | a3 | a4 | a5 | a6 | a7 | a8 | a9 | a10 |
|---|---|---|---|---|---|---|---|---|---|
| 1 | 1 | 1 | 1 | 1 | 1 | 1 | 1 | 1 | 1 |

| * | 1 | 2 | 3 | 4 | 5 | 6 | 7 |
|---|---|---|---|---|---|---|---|
| 1 | 6 | 5 | 2 | 3 | 4 | 1 | 7 |
| 2 | 4 | 3 | 7 | 1 | 2 | 6 | 5 |
| 3 | 7 | 6 | 3 | 4 | 5 | 2 | 1 |
| 4 | 1 | 7 | 4 | 5 | 6 | 3 | 2 |
| 5 | 2 | 1 | 5 | 6 | 7 | 4 | 3 |
| 6 | 3 | 2 | 6 | 7 | 1 | 5 | 4 |
| 7 | 5 | 4 | 1 | 2 | 3 | 7 | 6 |

| b1 | b2 | b3 | b4 | b5 | b6 | b7 | b8 | b9 | b10 |
|---|---|---|---|---|---|---|---|---|---|
| 4 | 1 | 6 | 3 | 7 | 5 | 2 | 4 | 1 | 6 |

| * | 1 | 2 | 3 | 4 | 5 | 6 | 7 | 8 | 9 |
|---|---|---|---|---|---|---|---|---|---|
| 1 | 8 | 7 | 2 | 3 | 4 | 5 | 6 | 1 | 9 |
| 2 | 6 | 5 | 9 | 1 | 2 | 3 | 4 | 8 | 7 |
| 3 | 9 | 8 | 3 | 4 | 5 | 6 | 7 | 2 | 1 |
| 4 | 1 | 9 | 4 | 5 | 6 | 7 | 8 | 3 | 2 |
| 5 | 2 | 1 | 5 | 6 | 7 | 8 | 9 | 4 | 3 |
| 6 | 3 | 2 | 6 | 7 | 8 | 9 | 1 | 5 | 4 |
| 7 | 4 | 3 | 7 | 8 | 9 | 1 | 2 | 6 | 5 |
| 8 | 5 | 4 | 8 | 9 | 1 | 2 | 3 | 7 | 6 |
| 9 | 7 | 6 | 1 | 2 | 3 | 4 | 5 | 9 | 8 |

| c1 | c2 | c3 | c4 | c5 | c6 | c7 | c8 | c9 | c10 |
|---|---|---|---|---|---|---|---|---|---|
| 1 | 8 | 2 | 9 | 5 | 7 | 3 | 4 | 1 | 5 |

Figure 3: Encryption using 2 different groups of different order

The Multi Level Indexed encryptor is represented as

$$QE^{I_r, I_s}_{h_1, h_2, \ldots h_n}(a_1, a_2, a_3 \ldots a_n) = e_1, e_2, e_3, \ldots e_n \qquad (7)$$

where $a_1, a_2, a_3 \ldots a_n$ is the input data and $e_1, e_2, e_3, \ldots e_n$ is the output vector $I_r$ and $I_s$ are called indices that are arrays which have the indices of quasigroups having corresponding order. The vector $(h_1, h_2, h_3 \ldots h_n)$ is the Hidden key or the Secret key. It is the output of the MEG-1 algorithm.

## 3.2 Decryption

This process is largely similar to the process of encryption which has just been discussed. The main point to note is the generation of the inverse matrix. The left inverse '\' is used for the quasigroup decryption as described in the Figure 4. The fundamental equations for encryption is:

$$D(a_1, a_2, a_3, \ldots a_n) = e_1, e_2, e_3, \ldots e_n \qquad (8)$$

where

$e1 = a \backslash a1$ and $ei = ai\text{-}1 \backslash ai$

To perform the process of decryption we need to first generate the inverse matrix of a given quasigroup and execute mapping procedure as described in the previous section (Figure 2), but we need to use equation (7) instead of (4). The decryptor for a multilevel indexed based algorithm may be defined as follows:

$$QD^{I_r,I_s}_{h_n,h_{n-1},..h_1}(e_1,e_2,e_3,...e_n) = a_1,a_2,a_3,...a_n \qquad (9)$$

The concept underlying the MLQE decryptor is lot similar to the MLQE encryptor that will be described in the next section.

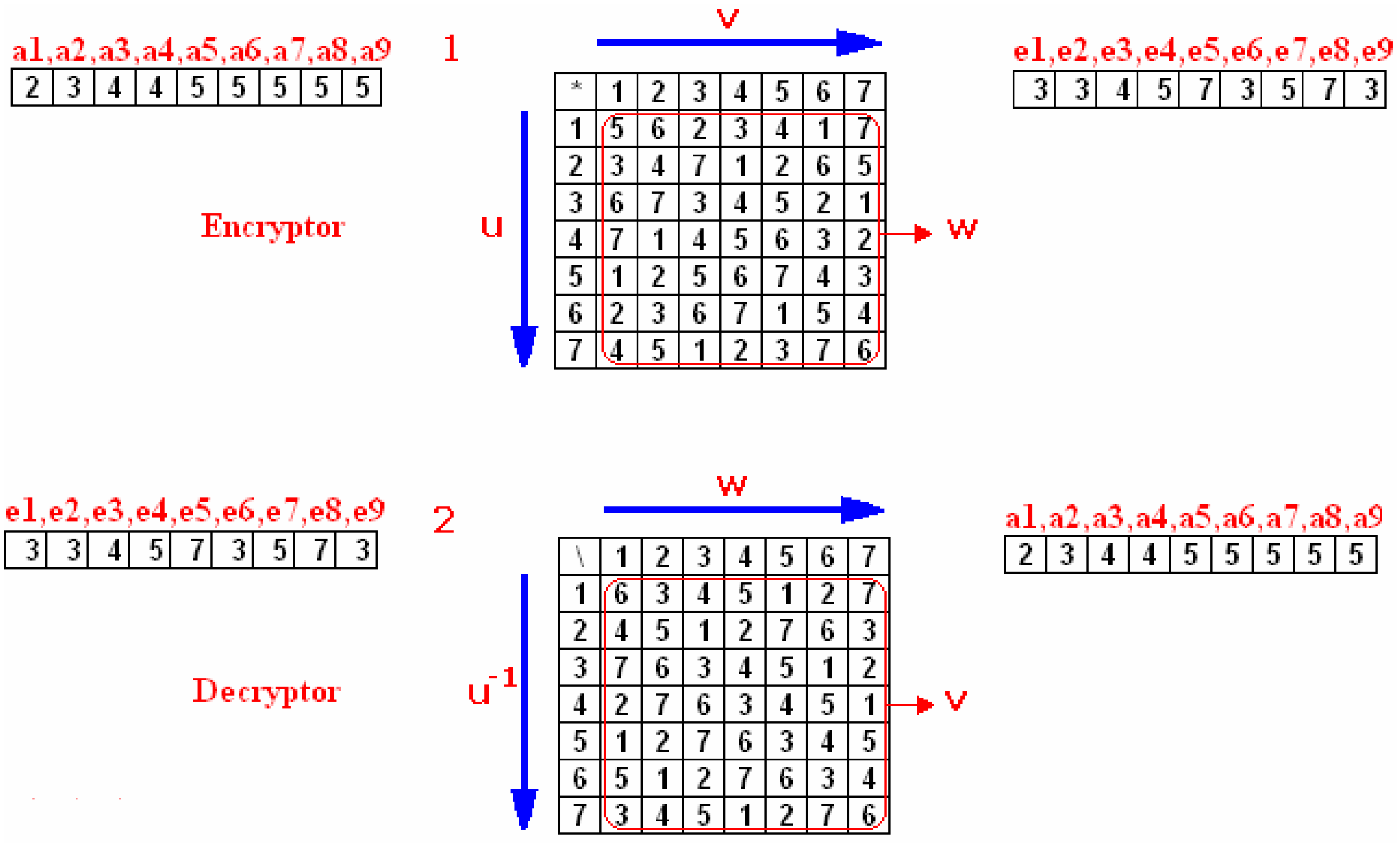

Figure 4: Determination of left division and the compete process of encryption and decryption

Some observations from the above illustration:

1. In Figure 4 the elements in the quasigroup (marked as 1) are labeled as *w*, the indices along the horizontal are labeled *v* and the indices along the vertical are labeled *u*.
2. The elements of the inverse (left-inverse) of quasigroup (labeled as 2) are labeled as *v* the indices along the horizontal are labeled *w* and the indices along the vertical are labeled $u^{-1}$ .

## 3.3. How to determine the left inverse of a given quasigroup

By executing the following algorithm, one can generate the left inverse of a given quasigroup:

1. Scan the row of the quasigroup

2. Locate an element *i* (start the value of *i* from 1) and note the value of *v* in the corresponding location of the inverse matrix.
3. Increment *i*
4. Go to step 1
5. This process happens till all the elements in the row are exhausted
6. Then go to the next row and repeat steps 1 through 3

## 4. Multi-Level-Index-Based-Quasigroup-Encryptor Receiver

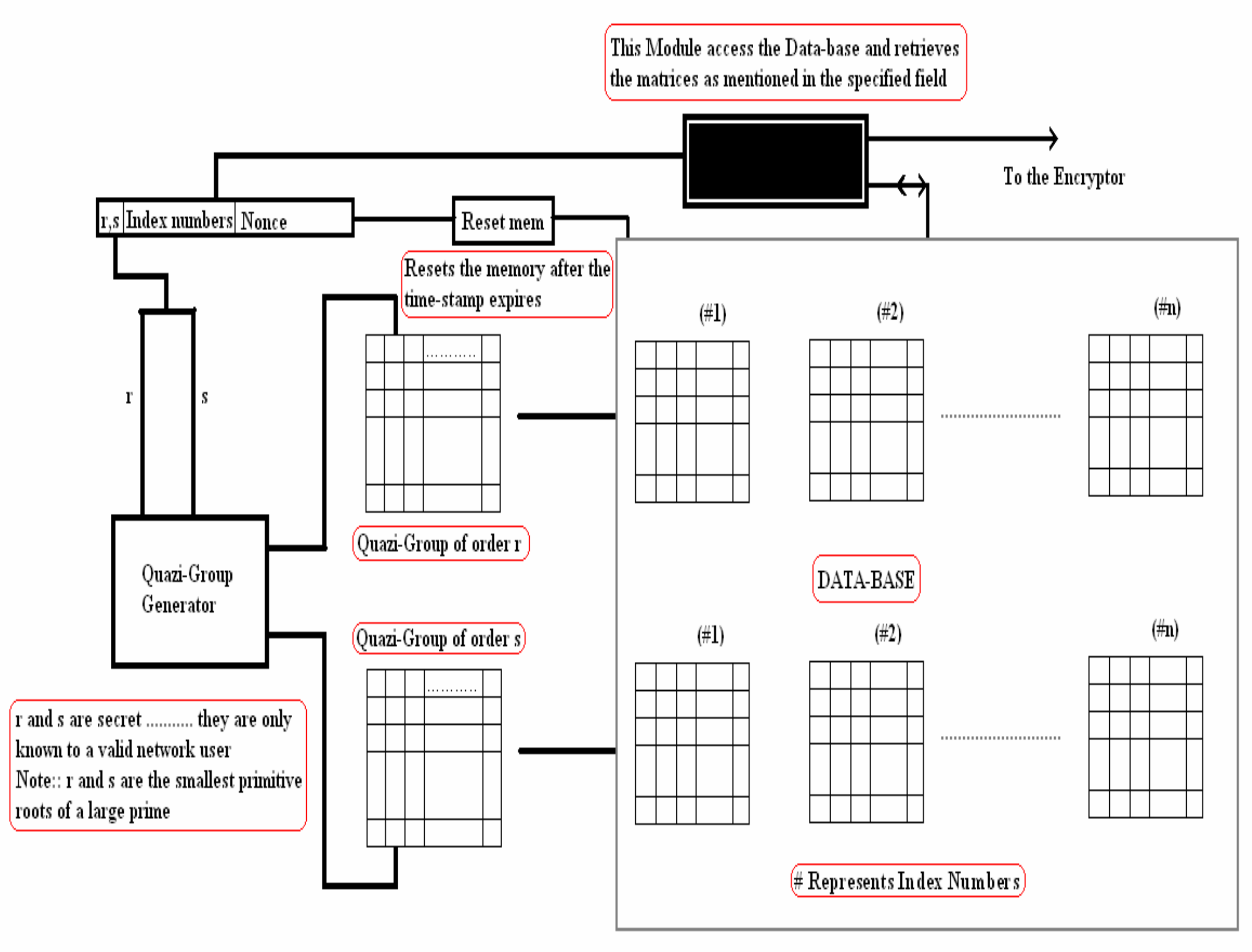

Figure 5: The MLQE Encryptor

The MLQE is a randomized algorithm that works on the encrypted key sent by a trusted authority. As illustrated in the above Figure 5, the input to this algorithm is a packet of data that consists of the following fields:

1. The first field consists of the orders of the two quasigroups that have to be generated
2. The second field consists of the Index numbers $I_1, \ldots I_r$ & $I_1 \ldots I_s$
3. The third field consists of the Nonce

The algorithm first generates a quasigroup of order r with all the required properties, followed by the generation of all the possible isotopes of that particular base quasigroup. These isotopes are stored in the data base in a specified manor. Later it generates another quasigroup of order *s* and stores the isotopes in a similar fashion.

These quasigroups may be accessed by the encryptor module based upon the index numbers supplied to it. The encryptor module codes the incoming data based upon the indices supplied to it, this module is also dependent on the inputs from another module that generates the multiplier seeds *q1…qn*. Experimental results have shown that this algorithm has the capability of scrambling data over a wide range depending upon the index numbers, *r, s*, and the multiplier elements.

## 5. THE PARAMETERS

In MLQE the final output data is dependent on the following factors:

1. Index numbers
2. Order of the index numbers
3. Order of the matrices *r* and *s*
4. The multiplier elements $(q_1, q_2, q_3, ... q_n)$

### 5.1 Index Numbers and their Order

This is perhaps the most important concept underlying the principle of working of a MLQE. The index numbers that can be generated may wary depending on the Nonce and the order of the matrices. The Nonce plays a major role in indexing, it has two important functions:

1. It is used by the receiver to reset the database after the Nonce expires.
2. It is used by the Trusted Authority to generate an index and the order of the matrices. This way even the trusted authority does not know the next key that is to be transmitted at any given point of time. The algorithm that generates the index numbers and the orders of the matrices is named FG-1 (for ‘frame generator’).

**Example 1.** Let us assume that the FG1 algorithm generates the values of *r, s* and the index numbers as shown in the example below. Let us assume the following values:

Let r = 150 and s = 270
Let $I_1, I_2, I_3, I_4, I_5, I_6$ = 1, 2, 3, 5, 4, 6
Let the Hidden key be: 2, 3, 5, 6, 1, 4
Let $d_1, d_2, d_3, d_4, d_5, d_6, d_7, d_8, d_9, d_{10}$ = 1,1,1,1,1,1,2,2,2,2 (Note: this example would show the true scrambling capability of the encryptor.)

The output data is the vector: $e_1, e_2, e_3 \ldots e_{10}$

It must be noted that in the example being considered, we encode the data using six different matrices and the final output data would have any number in between 1 and 270 (which the order of the second matrix). One might argue that the largest number in the encrypted sequence be considered as the order of the second matrix, but this is not correct because the encrypted data is produced by mapping the input data through several levels and the probability of the occurrence of the largest number becomes a function of the index vector which is in turn random. The simulation of the algorithm gives the following results:

| *Serial Num.* | *Index Numbers* | *Output Sequence* ($e_1,e_2,e_3,\ldots e_{10}$) |
|---|---|---|
| 1 | 1,2,3,5,4,6 | 126,95,162,16,123,93,163,216,176,249 |
| 2 | 1,6,5,4,2,3 | 125,92,154,267,83,17,30,268,107,14 |
| 3 | 1,4,2,5,6,3 | 126,95,162,16,123,93,163,216,176,249 |
| 4 | 1,5,6,4,2,3 | 126,95,162,16,123,93,163,216,176,249 |
| 5 | 1,4,5,2,6,3 | 127,99,172,36,158,149,247,66,71,199 |
| 6 | 1,4,6,2,5,3 | 121,82,137,244,58,268,30,36,209,214 |
| 7 | 2,3,5,6,1,4 | 270,255,190,9,140,21,91,27,67,169 |
| 8 | 5,6,2,3,1,4 | 123,84,136,240,62,45,48,22,113,252 |
| 9 | 2,5,3,1,6,4 | 1,255,189,7,137,17,86,21,60,161 |
| 10 | 2,3,5,1,6,4 | 1,255,189,7,137,17,86,21,60,161 |
| 11 | 2,3,5,4,6,1 | 4,261,200,24,161,49,127,72,122,235 |
| 12 | 1,1,1,1,1,1 | 124,87,139,232,13,161,90,208,152,109 |
| 13 | 2,2,2,2,2,2 | 2,255,193,18,159,55,146,110,186,63 |
| 14 | 3,3,3,3,3,3 | 5,259,180,220,231,134,164,31,237,149 |
| 15 | 4,4,4,4,4,4 | 269,240,131,104,126,62,65,3,113,2 |

From the above table the following deductions may be made:

1. The output sequence tends to be similar for some sequences of the index numbers. But it does not exhibit any periodicity and is unpredictable when we consider the practical case where the output ant any instance is dependent on other constraints such as the order of the matrix involved and the Hidden key.
2. One might argue that they can crack this code by simply doing a known plane text attack. But that is impossible because the algorithm that generates the keys has a random input (Nonce).

## 5.2. Order of the matrices r and s

In general there is no specified restriction on the order of the matrices supplied. However it is mandatory that the order of *r* must be smaller than that of *s*. One must also note that the final output sequence depends on the relative difference in the orders of the two matrices. It may be observed that the range of permissible sequence only depends upon the relative difference in between the order of the first and the order of the second matrix.

## 5.3. The Multiplier elements

The multiplier elements act like a seed to trigger the encryption process. In order to ensure the reliability of the decrypted data, the whole network must have this key (this is the reason for the terminology “Hidden Key”). This problem is resolved by embedding the special procedure MEG1 into all the valid network devices including the trusted Authority and the user hand sets.
The multiplier elements are depending upon the order of the matrix under consideration, the index number, and the Nonce. There are as many multipliers as index numbers. Thus this algorithm ensures the following:

1. It becomes virtually impossible to break this cipher based on known plaintext attack
2. It reduces the effort to encrypt and decrypt the data for a legitimate user while the computational requirement required to break this cipher is tremendously high for an illegitimate user.

# 6. The MEG-1 and FG-1 Algorithms

## 6.1. The FG 1 algorithm

From our study of the MLQE it is quite evident that the security of the entire system depends solely on the unpredictability of the key (r, s, index elements and Nonce).The FG 1 module ensures that even the Trusted authority doesn’t know the key that it would be transmitting in the next instant (next time slot).

This algorithm generates the order of the matrices *(r,s)*, index numbers and the Nonce based on a random input from a pseudo-random number generator. This module is essential due to the following reason:

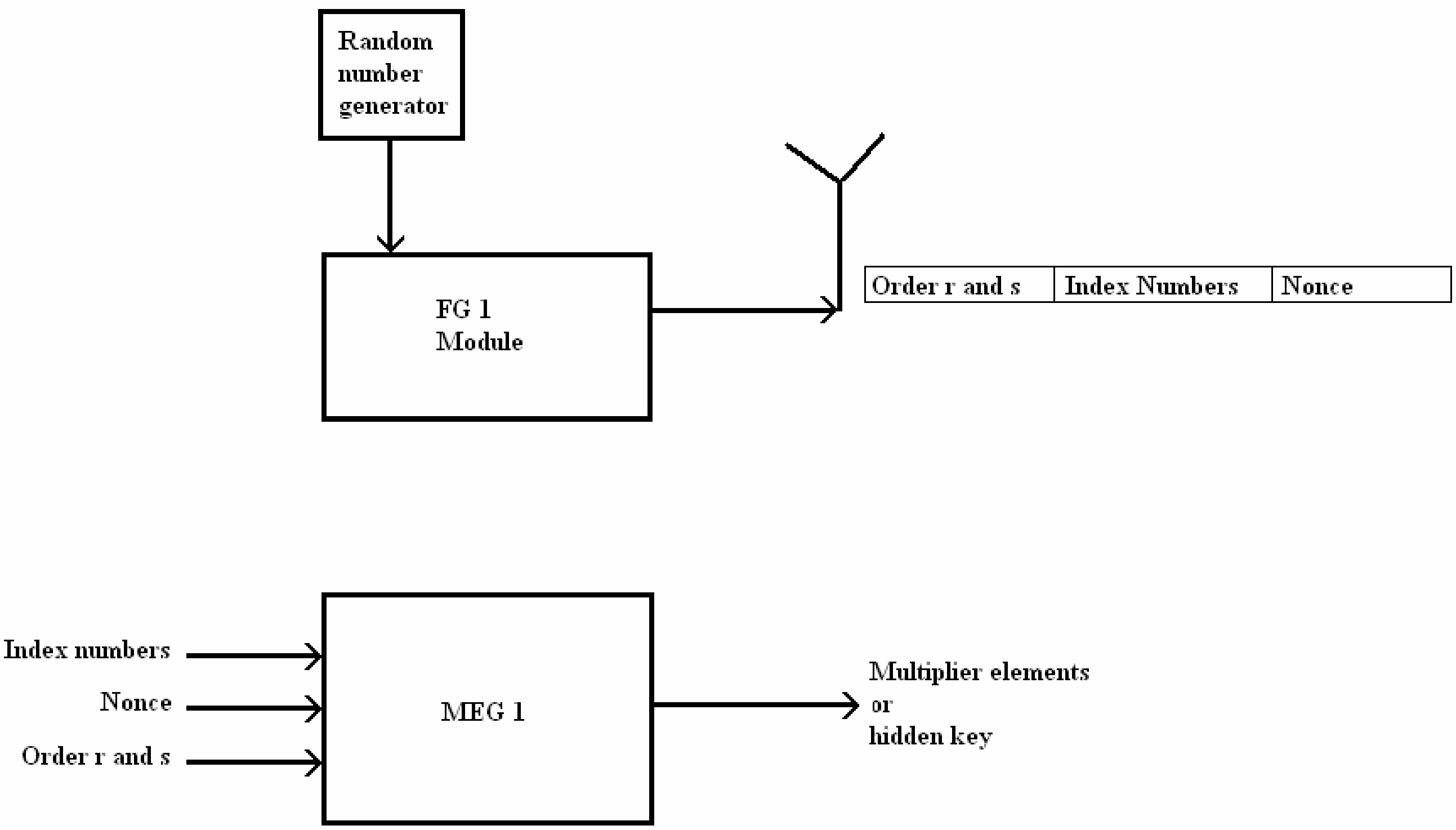

Figure 6: The MEG 1 and FG 1 algorithms

1) This module ensures that the next sequence of indices generated by the system is un-predictable.
2) The Nonce is it self dependent on the random input which makes the system more dynamic.
3) The security of the system is dependent on the period of the random number generator and the network time (network time is denoted by ‘T’. This time judges the value of Nonce‘t’ that is to be sent in the next frame and at any given instant the value of t should not exceed the value of ‘T’).
4) The security also depends on the other factors like length of the random symbol, number of indices and the order of the Quasi-groups.

## 6.2. The MEG 1 algorithm

While the FG 1 algorithm is only required at the trusted authority the MEG 1 algorithm plays a major role in the encryption and the decryption process. During the process of encryption, it generates the ‘Hidden key’ (consisting of the multiplier elements) and during the process of decryption the same key has to be generated at the receiver in order to ensure proper decryption. The MEG 1 algorithm takes the entire incoming frame and generates the multiplier elements.

### 6.2.1. Nonce *t*

The Nonce acts like a random input to this algorithm. Since the hidden key generated is dependent on the value of the Nonce supplied to it, this algorithm would generate different different “Hidden keys” even if the same orders (r,s) and index numbers are supplied to it . The Nonce might be given by the formula

$$\text{Nonce } t = f\ (prsg) \qquad (10)$$

where *prsg* stands for *Pseudo Random Number Generator*. We can even use the same random number that is given as the input to the FG-1 module. Since the Nonce is generated by the FG 1, it is evident that it remains constant for the entire network till it expires and a new transmission is made.

There is a tradeoff of security and overhead involved here. It is required that this Nonce changes as frequently as possible, this however increases the computational costs of the entire network. In order to avoid any unnecessary overload the Nonce must be above a certain specified limit. Assume that this arbitrary limit is given by $T_1$:

$$T_1 < \text{Nonce ‘}t\text{’} < T \qquad (11)$$

The above equation can be very significant in determining the network overload. If $T_1$ is very small then the devices in the network would have to recalculate the groups and indices several times per session and on the other hand if $T_1$ is made too large ($T_1 \rightarrow T$) then there is a high chance of the network being compromised.

### 6.2.2. Order of r, s

In a quasigroup encryption the multiplier element for a certain level must belong to the quasigroup under consideration. This implies that the set of multipliers ($q_1,q_2,q_3\ldots qn$) belong to the set of quasigroups (corresponding indices) that are being used at that particular instant for the process of encryption.

Let us suppose that $I_1$, $I_2$, $I_3$ are the indices corresponding to the quasigroup of order *r* and let $I_4$, $I_5$, $I_6$ be the indices corresponding to the Quasigroup of order *s*. Let us suppose that the MEG 1 generates $q_1$, $q_2$, $q_3$, $q_4$, $q_5$, $q_6$. Now $q_1$, $q_2$, $q_3$ are used as the multiplier elements during the encryption using the quasigroup of order *r*. For the data to be properly mapped, $q_1$, $q_2$, $q_3$ must belong to the group of order *r*. Likewise, if the multipliers $q_4$, $q_5$, $q_6$ are used during the encryption using the second group then it is implied that they belong to that group.

## 7. Randomization measurements

In this section we will study the characteristics of the output data by supplying some test data to the MLQE. In general the text information sent can be very random, repetitive or have the certain characteristics (in English text, E has the largest probability) that would make the encoded file vulnerable to the known plain text attack. We consider several cases of highly redundant and random data. To compute the autocorrelation function, the mean of the input data is subtracted from the values, ensuring that the autocorrelation function has positive and negative values. The autocorrelation function is not normalized.

### 7.1. Case 1: A constant input

Let us suppose that the input data consists of a series of 'K' followed by space. (This example would demonstrate the scrambling nature of the MLQE).

Input data: K K K K K K K K K K K
Encoded data: 19 15 1 3 30 18 25 4 5 20 17 33 33 2 5 37 31 33 36 30 14 38
Key used: 35, 41, 5, 4, 2, 1, 6, 3

It may be observed that though the input sequence has only 2 values, the output is mapped to a range of values from '1' to '38'. It must be noted that the first 2 integers in the key are the order of the matrices and the remaining are the encryption indices.

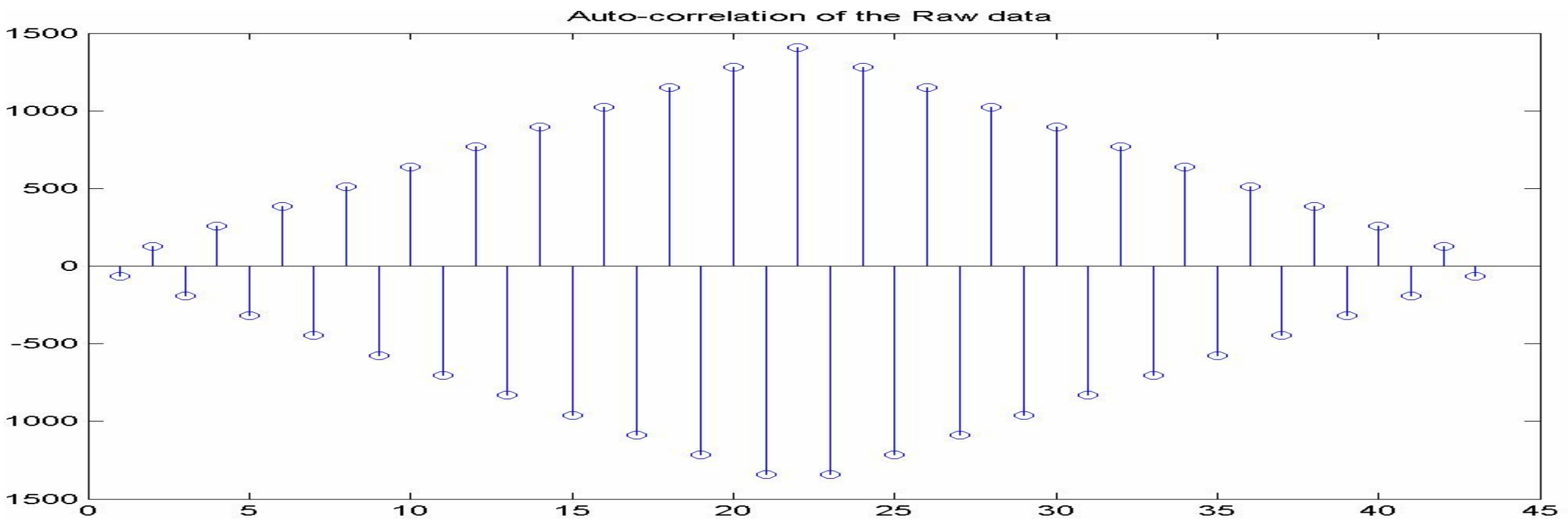

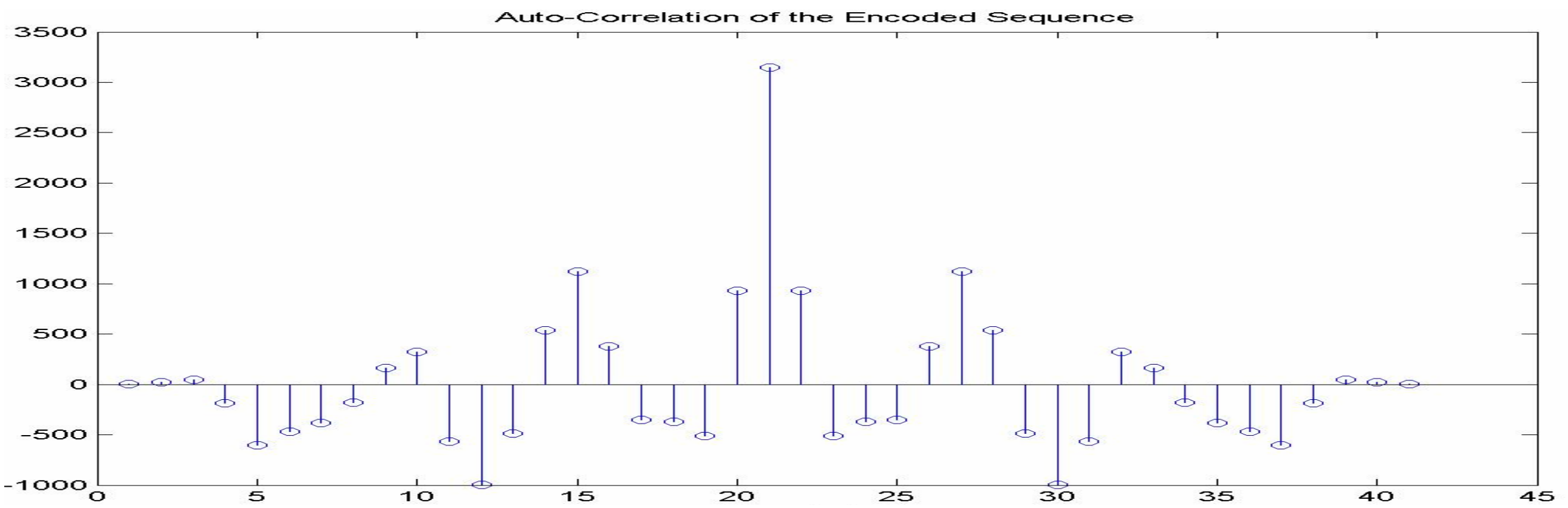

Figure 7: Auto-correlation of the raw data and the encoded data for Case 1

## 7.2. Case2: Plain text

In this case we consider a plain text data, here it can be observed that in the given text the Letter 'A' occurs 4 times however it is mapped to four different values (A1, A2, A3, A4 = 31, 19, 40, 21 where A1, A2, A3, A4 represent the four As that have occurred in the text "OOM NAMAH SHIVAYA")

Input data: OOM NAMAH SHIVAYA
Encoded data: 23 33 40 26 31 28 19 34 37 41 12 8 25 40 33 21 19
Key used: 35, 41, 5, 4, 2, 1, 6, 3

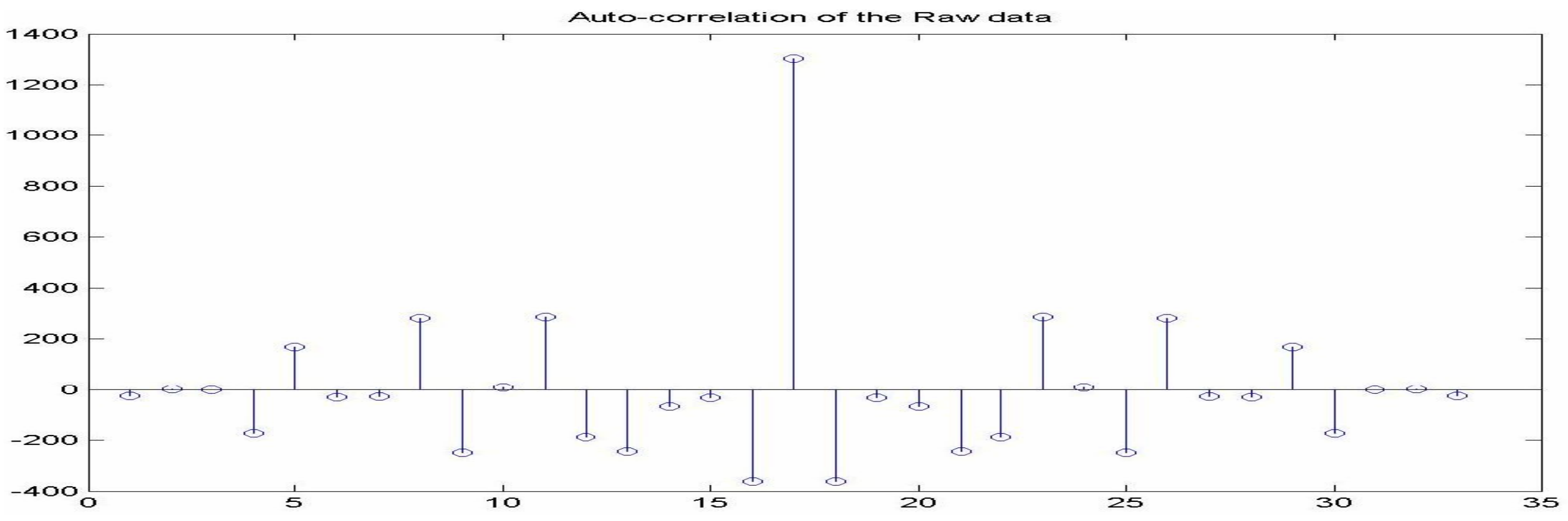

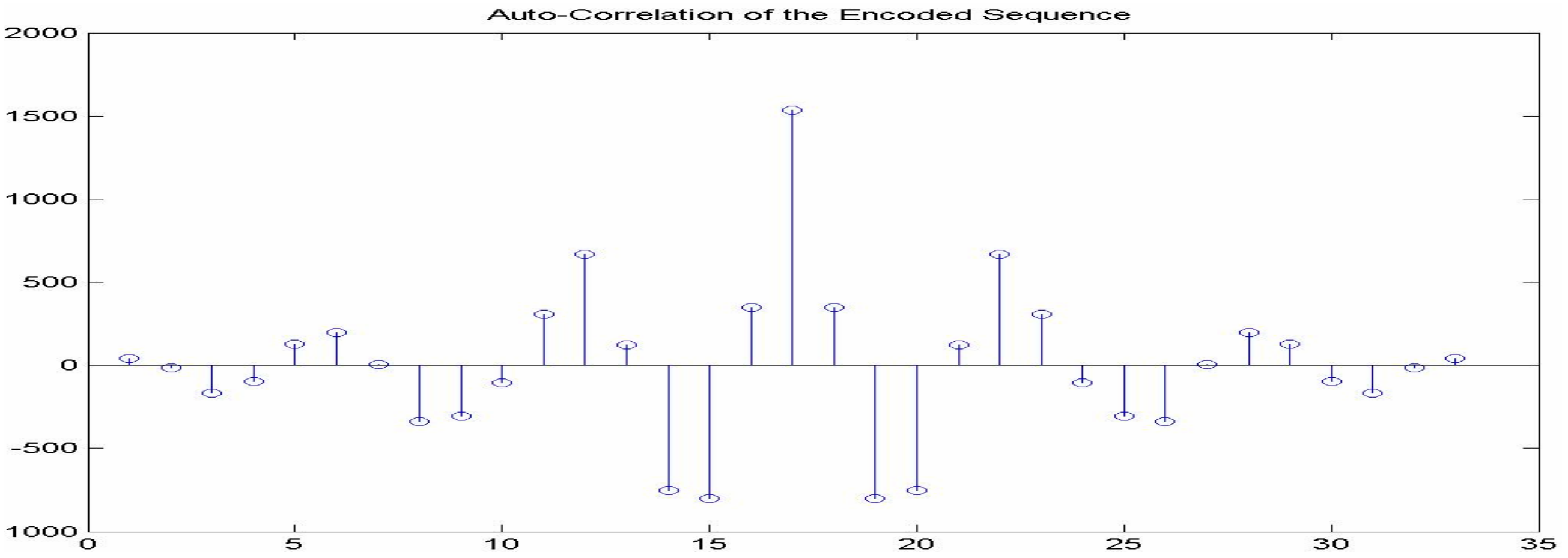

Figure 8: Auto-correlation of the raw data and the encoded data for Case 2

## 7.3. Case 3: Random input

In this case we test the response of the MLQE when the input data is itself random.

Input data: E M V C W J F A Z
Encoded data : 11 10 41 35 12 7 28 31 22 6 19 5 15 20 10 9 16
Key used: 35, 41, 5, 4, 2, 1, 6, 3

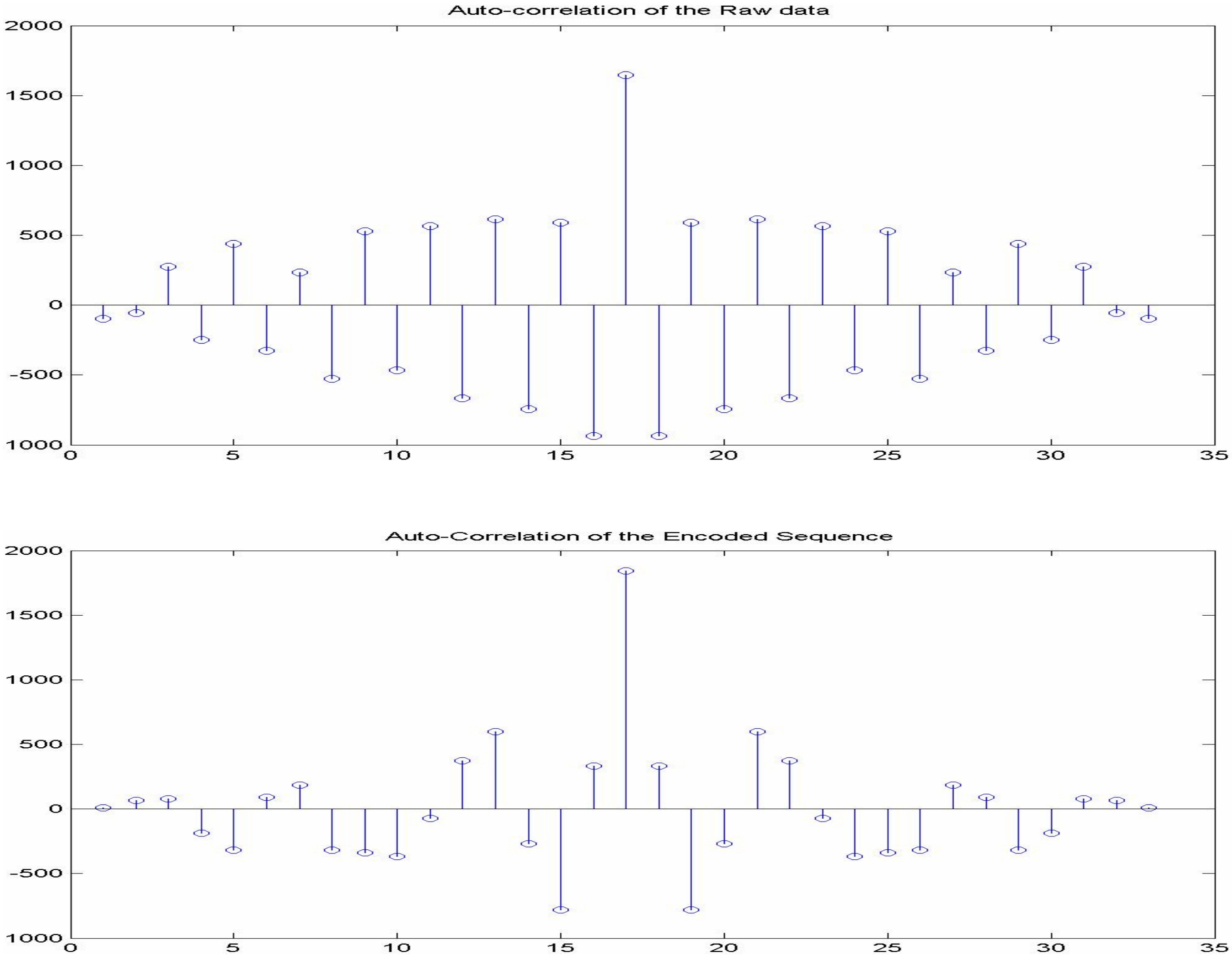

Figure 9: Auto-correlation of the raw data and the encoded data for Case 3

It may be observed from figures 7, 8, 9 that the output characteristics are a lot similar for all the three cases even though the inputs are very different.

## 7.4. Case 5: Long English text

Input data:

ONE DAY A COUNTRYMAN GOING TO THE NEST OF HIS GOOSE FOUND THERE AN EGG ALL YELLOW AND GLITTERING WHEN HE TOOK IT UP IT WAS AS HEAVY AS LEAD AND HE WAS GOING TO THROW IT AWAY BECAUSE HE THOUGHT A TRICK HAD BEEN PLAYED UPON HIM. BUT HE TOOK IT HOME ON SECOND THOUGHTS AND SOON FOUND TO

Encoded data:

23 32 17 38 35 8 12 21 24 39 1 26 6 39 15 1 3 6 39 34 4 11 40 29 4 19 20 2 9 34 20 5 6 29 37 6 5 41 23 17 29 37 16 30 28 33 13 41 30 18 11 40 40 19 40 13 23 21

30 34 27 40 8 37 8 14 5 15 23 20 30 32 35 24 26 34 21 39 41 14 4 5 16 26 37 33
11 3 37 34 8 12 33 16 18 36 2 26 12 29 22 36 27 16 18 25 17 19 8 8 33 6 28 26
17 39 19 33 22 38 34 32 27 12 4 19 5 29 16 33 33 39 34 37 38 36 24 25 5 14 39 3
38 13 20 30 30 26 37 31 17 8 3 19 8 33 17 34 19 27 25 6 19 30 19 19 7 10 1 30 20
3 29 6 40 33 19 13 36 39 35 25 30 22 34 16 18 13 31 19 28 4 28 9 39 14 24 4 6
22 10 39 32 21 15 37 19 19 4 29 13 26 4 16 6 29 18 29 7 21 19 19 16 38 19 9 37
6 16 14 25 22 26 14 19 34 11 13 3 13 27 27 17 9 17 34 30 3 9 15 34 15 3 11 22
18 39 26 19 41 32 29 10 20 11 11 13 35 40 39 30 24 10 12 17 3 28 41 26 5 32 19 19

Key used: 35, 41, 5, 4, 2, 1, 6, 3

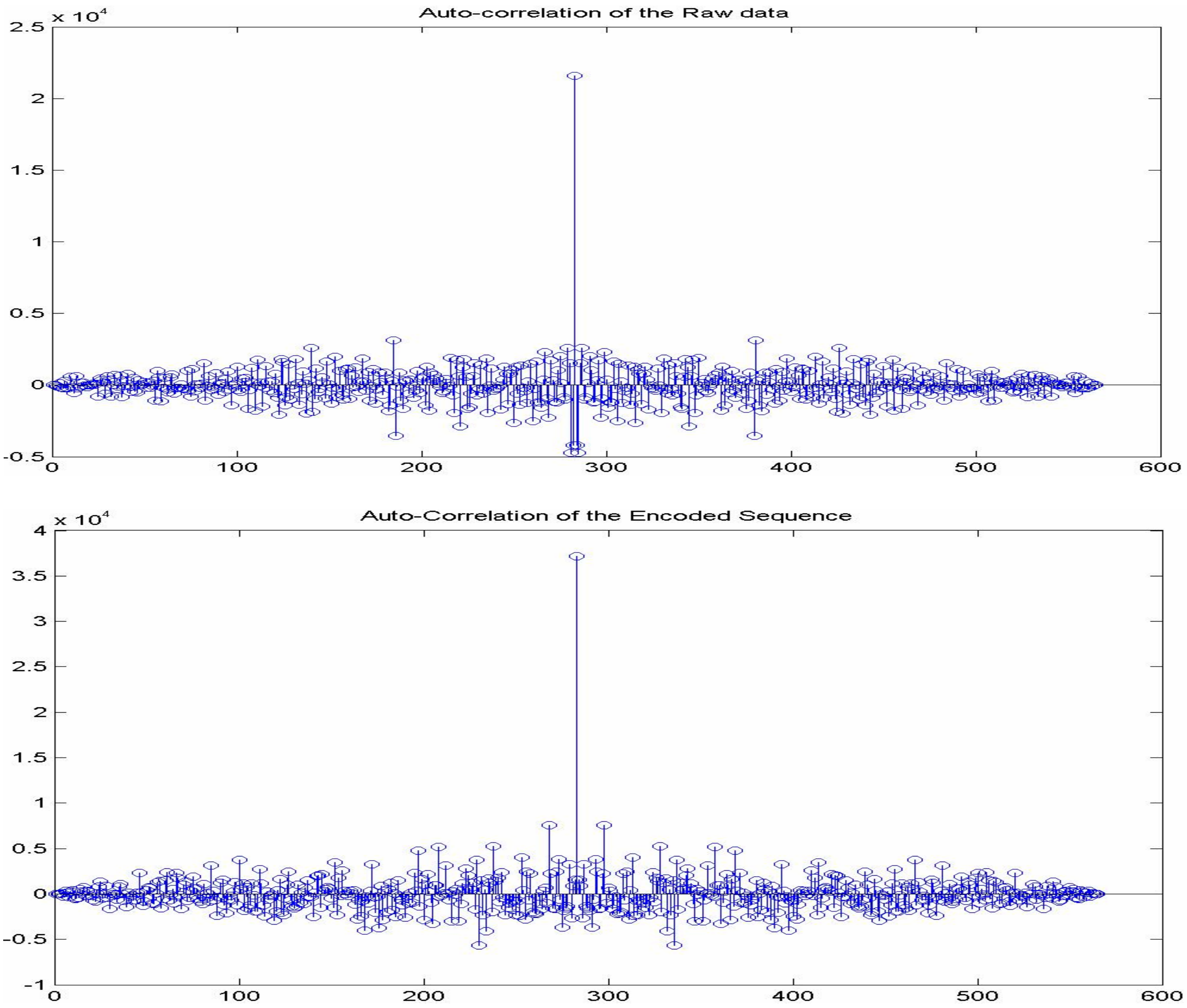

Figure 10: Auto-correlation of the input data and the encoded data for Case 5

## 7.5. Case 6: Long constant text

Input data:

E E E E E E E E E E E E E E E E E E E E E E E E E E E E E E E E E E E E E E E E E E E E E E E
E E E E E E E E E E E E E E E E E E E E E E E E E E E E E E E E E E E E E E E E E E E E E E E
E E E E E E E E E E E E E E E E E E E E E E E E E E E E E E E E E E E E E E E E E E E E E E E
E E E E E E E E E

Encoded data:

11 10 24 28 33 11 23 8 28 10 17 20 2 40 29 15 22 2 30 6 30 7 29 37 22 10 7 9 18
4 29 38 12 30 17 33 7 10 27 28 38 38 33 12 11 19 4 12 29 6 7 12 9 37 16 12 34 26
18 5 40 17 39 36 41 38 36 16 31 34 8 32 35 9 33 1 34 29 20 1 33 7 33 34 33 13 9
5 38 3 39 38 6 32 18 34 16 21 34 9 41 24 12 34 30 35 19 7 32 7 32 1 24 21 20 5
27 11 23 7 12 27 35 12 9 18 2 33 41 9 33 32 23 35 5 29 15 16 23 5 21 40 3 6 16
4 27 18 23 12 30 1 27 15 19 17 22 18 35 23 38 27 36 18 35 24 17 27 6 9 23 37 32
36 23 41 6 5 28 20 26 2 6 6 33 32 8 21 3 12 1 32 29 22 6 26 21 4 33 7 38 33 16
23 5 8 12 21 19 40 6 24 1 28 9 4 5 7 37 2 35 8 23 4 16 7 3 12 22 18 36 27 31
37 19 34 35 7 17 8 36 9 34 21 30 28 7 27 40 35 25 16 19 10 15 20 6 8 18 1 22 8
34 22 25 36 19 32 40 29 9 35 34 10 35 41 34 35 11 17 23 11 36 35 22 25 34 38 14
25 39 12 34 27 26 35 26 20 27 1

Key used: 35, 41, 5, 4, 2, 1, 6, 3

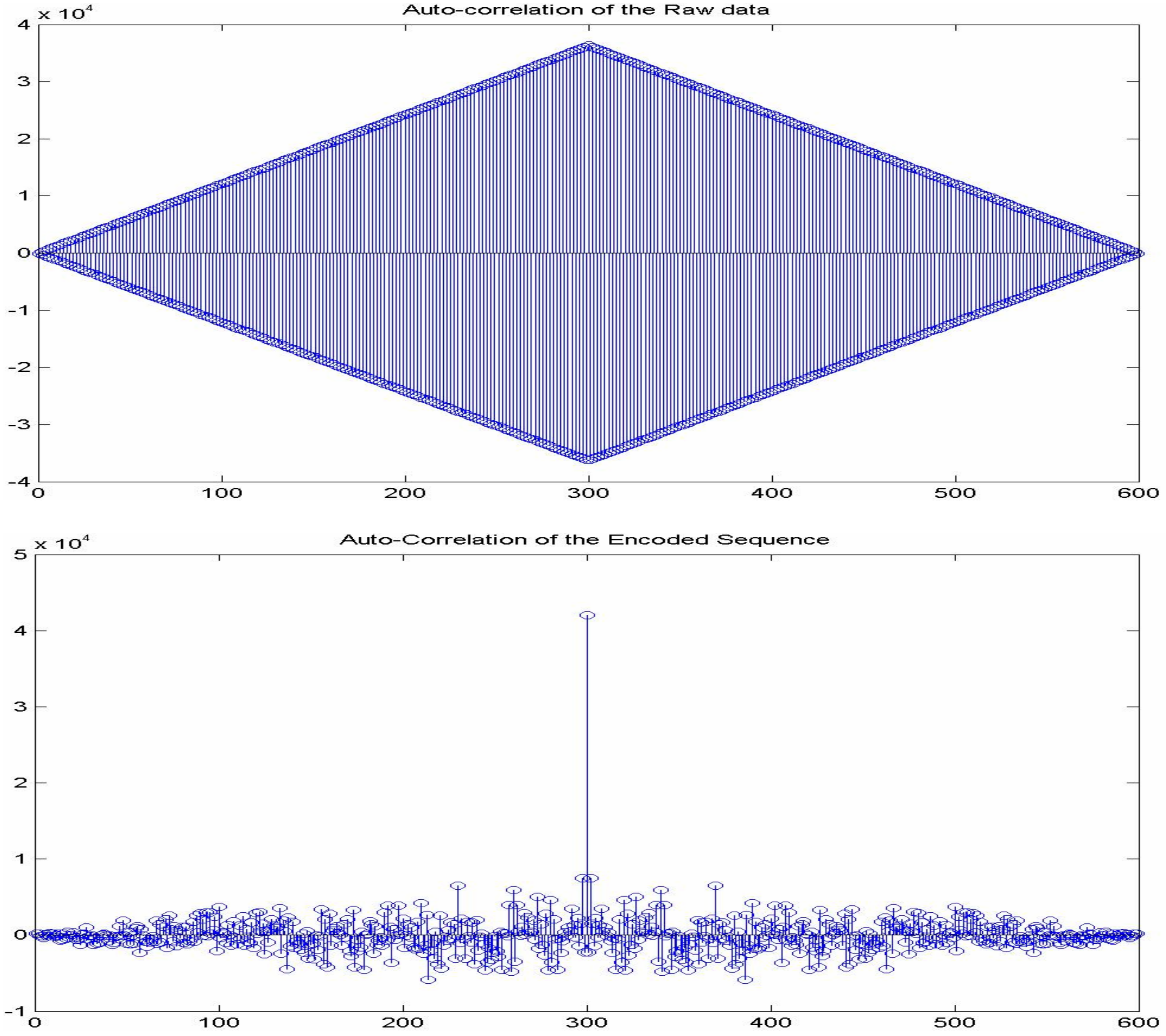

Figure 11 Auto-correlation of the raw data and the encoded data for Case 6

We can see from the autocorrelation graphs that the encrypted sequence is essentially two-valued and, therefore, very random, even in cases where the input was constant, as in Figure 11.

## 8. Concluding Remarks

We have shown in this article that quasigroup scrambling constitutes an excellent method of encryption and generation of pseudo-random sequences. The randomization obtained is very good, and, therefore, there would be many practical applications of symmetric cryptography where this method can find use. Apart from text scrambling, we propose application of it to speech scrambling systems.